\documentclass[conference]{IEEEtran}
\IEEEoverridecommandlockouts
\usepackage{cite}
\usepackage{amsmath,amssymb,amsfonts}
\usepackage{graphicx, color}
\usepackage{subcaption}
\usepackage[export]{adjustbox}
\usepackage{textcomp}
\usepackage{url} 
\usepackage{makecell}
\usepackage[table]{xcolor} 
\usepackage{paralist}
\usepackage{enumitem}
\usepackage{multirow}
\usepackage{hyperref}
\usepackage{algorithmic}
\usepackage{xcolor}

\usepackage{enumitem}

\usepackage[linesnumbered,ruled,vlined]{algorithm2e}
\hypersetup{
    colorlinks=true,
    linkcolor=red, 
    citecolor=blue, 
    urlcolor=magenta    
}

\def\BibTeX{{
    \rm B\kern-.05em{\sc i\kern-.025em b}\kern-.08em
    T\kern-.1667em\lower.7ex\hbox{E}\kern-.125emX}}

\newcommand{\smallurl}[1]{\footnotesize\url{#1}}

\definecolor{baselinecolor}{gray}{.9}

\begin{document}

\title{UU-Mamba: Uncertainty-aware U-Mamba for Cardiac Image Segmentation}

\author{
    \IEEEauthorblockN{
        Ting Yu Tsai\textsuperscript{1},
        Li Lin\textsuperscript{2},  
        Shu Hu\textsuperscript{2},
        Ming-Ching Chang\textsuperscript{1}, 
        Hongtu Zhu\textsuperscript{3},
        Xin Wang\textsuperscript{1,*} \thanks{* Coressponding author.}
    }
    \IEEEauthorblockA{
        \textsuperscript{1}University at Albany, State University of New York, {\tt \small \{ttsai2, mchang2, xwang56\}@albany.edu}\\
        {\textsuperscript{2}Purdue University, {\tt \small \{lin1785,  hu968\}@purdue.edu} }\\
        \textsuperscript{3}University of North Carolina at Chapel Hill, {\tt \small 
        htzhu@email.unc.edu}\\
    }
}

\maketitle
\thispagestyle{plain}
\pagestyle{plain}

\begin{abstract}
Biomedical image segmentation is critical for accurate identification and analysis of anatomical structures in medical imaging, particularly in cardiac MRI. Manual segmentation is labor-intensive, time-consuming, and prone to errors, highlighting the need for automated methods. However, current machine learning approaches face challenges like overfitting and data demands. To tackle these issues, we propose a new UU-Mamba model, integrating the U-Mamba model with the Sharpness-Aware Minimization (SAM) optimizer and an uncertainty-aware loss function. SAM enhances generalization by locating flat minima in the loss landscape, thus reducing overfitting. The uncertainty-aware loss combines region-based, distribution-based, and pixel-based loss designs to improve segmentation accuracy and robustness. Evaluation of our method is performed on the ACDC cardiac dataset, outperforming state-of-the-art models including TransUNet, Swin-Unet, nnUNet, and nnFormer. Our approach achieves Dice Similarity Coefficient (DSC) and Mean Squared Error (MSE) scores, demonstrating its effectiveness in cardiac MRI segmentation. 
The code can be accessed at \smallurl{https://github.com/tiffany9056/UU-Mamba}.
\end{abstract}

\section{Introduction}
Biomedical image segmentation is essential in medical imagery analysis for accurately identifying and delineating anatomical structures and anomalies. In cardiac imaging, segmenting the heart and its components from Magnetic Resonance Imaging (MRI) is critical for diagnosing cardiovascular disorders, the development of treatment plans, and assessing therapeutic outcomes \cite{bernard2018deep, litjens2017survey}. Cardiac MRI provides high-quality images that detail the structure, function, and composition of the heart \cite{fahmy2019}. However, manual segmentation of the cardiac structures from these images is labor-intensive, time-consuming, and prone to observer variability, highlighting the need for automated segmentation techniques to improve efficiency and consistency \cite{petitjean2011review, maier2019gentle}.

Cardiac MRI segmentation is crucial but challenging due to variations in heart anatomy, pathological changes, and imaging artifacts. Conventional methods like thresholding and edge detection often fail to capture the complexities of cardiac structures. Machine learning techniques, particularly convolutional neural networks (CNNs) and other Deep Learning models, have shown promise by learning intricate patterns from large datasets. However, these models require extensive annotated data, significant computational resources, and may struggle to generalize across different patient groups and imaging settings. Current research aims to develop more robust, efficient, and generalizable segmentation algorithms to enhance the accuracy and reliability of cardiac MRI analysis~\cite{litjens2017survey}.

\begin{figure}[t]
\centerline{\includegraphics[width=0.95\linewidth]{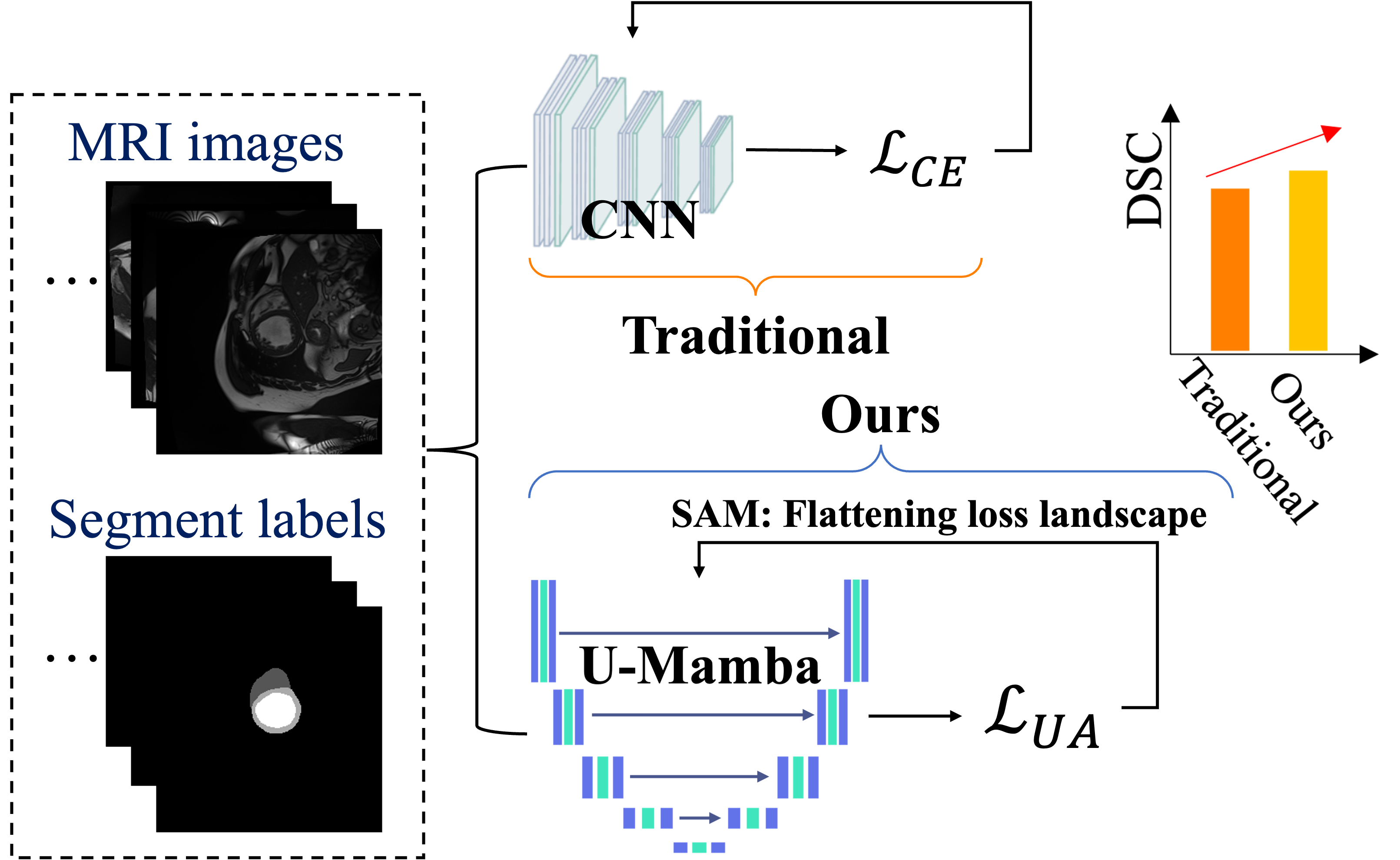}}
    \caption{
    Comparison between our method and traditional approach. Traditionally, a deep learning model is trained using the cross-entropy loss $\mathcal{L}_{CE}$.
    Our method enhances U-Mamba by unitizing Uncertainty-aware loss $\mathcal{L}_{UA}$, which is optimized via the SAM optimizer over a flattened loss landscape.
    }
    \vspace{-3mm}
    \label{fig:introduction}
\end{figure}

In this paper, we address these challenges by proposing the UU-Mamba model, an extension of the U-Mamba model~\cite{ma2024umamba}, and improve the training of it by incorporating a {\bf novel uncertainty-aware loss} and a Sharpness-Aware Minimization (SAM) optimizer. Specifically, our newly introduced uncertainty-aware loss function integrates three components:
\begin{enumerate}[leftmargin=16pt] \itemsep -.1em
    \item {\em Region-based loss}: Used for object detection and localization ({\em e.g.}, Dice coefficient loss).
    \item {\em Distribution-based loss}: Compares predicted and ground truth distributions ({\em e.g.}, cross-entropy loss). 
    \item {\em Pixel-based loss}: Measures differences at the pixel level ({\em e.g.}, focal loss).
\end{enumerate}

By employing {\bf auto-learnable weights}~\cite{cipolla2018} rather than fixed weights, our model can dynamically adjust the contribution of each loss component based on the uncertainty associated with each prediction. This approach allows the model to prioritize confident predictions and reduce the impact of ambiguous or noisy data. The auto-learnable weights enable the model to balance different aspects of the segmentation task adaptively, improving overall performance and robustness, and addressing issues like class imbalance\cite{azad2023}. 

To improve the generalization of our model, we utilize the {\bf Sharpness-Aware Minimization (SAM) optimizer}, which finds parameter values that result in flat minima in the loss landscape. This technique improves the model's ability to generalize~\cite{foret2021sharpness,lin2024robust,lin2024robust2,lin2024robust3,lin2024preserving}. and mitigates overfitting, a common issue in deep learning applications for medical imaging~\cite{Mariam2024}. 

Fig.~\ref{fig:introduction} compares our method against other existing methods.
In summary, we make the following key contributions:
\begin{itemize}[leftmargin=10pt] \itemsep -.1em
    \item We introduce an uncertainty-aware loss combining region-based loss ({\em e.g.}, Dice coefficient loss), distribution-based loss ({\em e.g.}, cross-entropy loss), and pixel-based loss ({\em e.g.}, focal loss). This design allows the model to prioritize confident predictions and reduce the influence of noisy data.
    \item We integrate Sharpness-Aware Minimization (SAM) optimization into our model. SAM optimization find parameter values that lead to flat minima in the loss landscape, which mitigates overfitting and reduces sensitivity to small perturbations in the input data.
    \item We evaluate and compare our method against state-of-the-art models (TransUNet~\cite{Chen2021TransUNet}, Swin-Unet~\cite{Cao2021SwinUnet}, nnUNet~\cite{Isensee2021nnUNet}, U-Mamba~\cite{ma2024umamba}),  demonstrating superior performance, achieving the highest Dice Similarity Coefficient (DSC) and lowest Mean Squared Error (MSE) by promoting sharper and more precise segmentation boundaries improving segmentation accuracy and robustness on the ACDC dataset.
\end{itemize}

\begin{figure*}[t]
    \centerline{
    \includegraphics[width=1\textwidth]{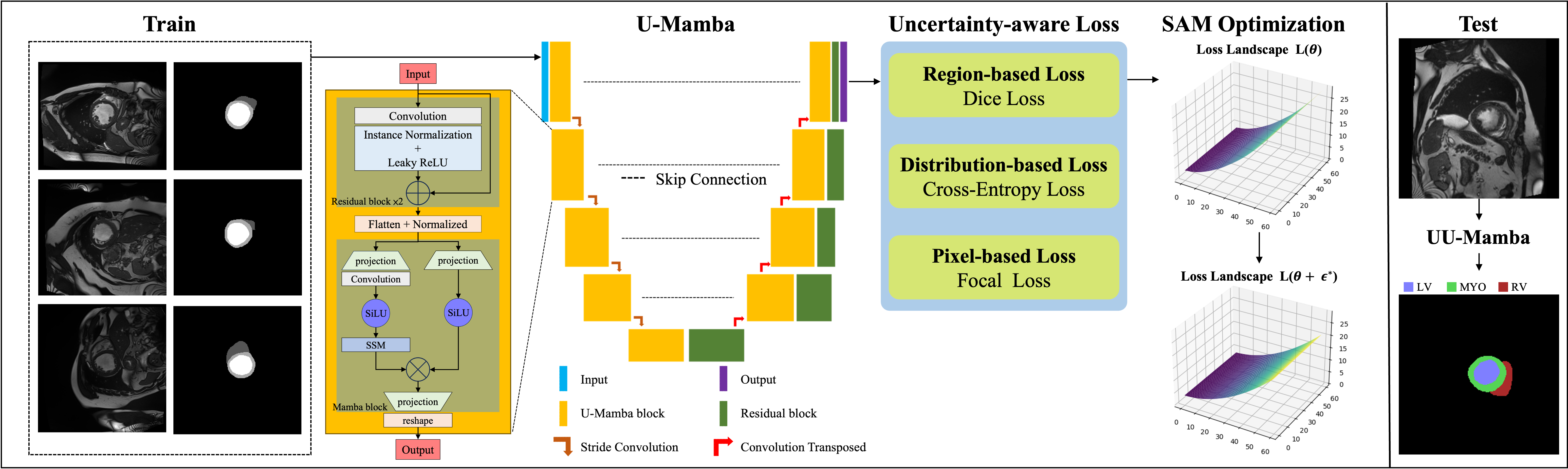}
    \vspace{-1mm}
    }
\caption{Overview of our proposed {\bf UU-Mamba model}: Leveraging the U-Mamba architecture, we encode input images and incorporate a novel uncertainty-aware loss function. Optimization is performed using the Sharpness-Aware Minimization (SAM) optimizer, which operates within a flattened loss landscape. Experiments on the ACDC dataset perform 3D heart segmentation on cardiac MRI images, delineating the right ventricle (RV), myocardium (Myo), and left ventricle (LV).
}
\vspace{-3mm}
\label{fig:Overview_proposed_model}
\end{figure*}
\section{Related Work}
\subsection{Cardiac MRI Segmentation}

Cardiac MRI segmentation has advanced significantly with deep learning methods, particularly Convolutional Neural Networks (CNNs) \cite{wang2024artificial}. 
While cardiac segmentation datasets such as ACDC have played a crucial role in this progress, existing methods still face challenges \cite{bernard2018deep}.
Bernard {\em et al.}~\cite{bernard2018deep} showed that CNNs, especially U-Net and its variants, achieve high accuracy in cardiac MRI segmentation. However, these methods often struggle with patient-specific variances and suffer from overfitting and class imbalance.
Isensee {\em et al.}~\cite{Isensee2021nnUNet} improved segmentation accuracy by combining U-Net and V-Net architectures, relying heavily on extensive annotations. Despite improvements, their approach still struggles to generalize to new data due to its dependence on large annotated datasets.
Transfer learning has been explored to mitigate the scarcity of labeled data.
Chen {\em et al.}~\cite{chen2020deep} pre-trained models on the ACDC dataset and fine-tuned them on smaller datasets. This approach improves performance but can suffer from domain shift issues when applied to different data distributions.
Traditional methods often use standard loss functions like Cross-Entropy (CE) loss, which may not handle class imbalance or the fine details required for accurate cardiac segmentation. These methods also tend to neglect optimizing for flatter minima in the loss landscape, which can enhance generalization.


\subsection{Mamba for Medical Image Segmentation}

The Mamba architecture~\cite{gu2023mamba} represents a significant advancement in medical image segmentation, combining the strength of CNNs and Vision Transformers (ViTs). It excels in managing long sequences and integrating global contextual information, which is crucial for precision in medical imaging tasks~\cite{wang2024mambaunet}. 
Building on this foundation, U-Mamba~\cite{ma2024umamba} enhances the traditional U-Net framework by incorporating attention mechanisms and multi-scale processing. This significantly improves segmentation accuracy and robustness by focusing on pertinent details within complex anatomical structures and effectively processing a range of scales, from broad contextual information to precise low-level details.
U-Mamba also integrates deep supervision to speed up training and improve convergence, making it efficient and reliable for clinical use, where fast and accurate image processing is crucial. The adaptability of the Mamba architecture is further demonstrated in specialized variants like Weak-Mamba-UNet~\cite{wang2024weakmambaunet}, which excels in scribble-based segmentation tasks, handling complex scenarios with enhanced performance. In summary, models based on U-Mamba facilitates superior segmentation across various medical applications, such as cardiac MRI and histopathological imaging.



\section{Method}
Fig.~\ref{fig:Overview_proposed_model} shows the overview of our proposed UU-Mamba architecture with training improvements. 
$\S$~\ref{sec:Mamba:UMamba} describes the Mamba block and the U-Mamba network, detailing the integration of state space models and their role in capturing long-range dependencies.
$\S$~\ref{sec:uncertainty-aware:loss} presents our proposed uncertainty-aware loss, explaining how it combines multiple loss functions to enhance model performance and robustness.
$\S$~\ref{sec:SAM:opt} describes the Sharpness-Aware Minimization Optimization, highlighting its benefits in achieving flat minima in the loss landscape to improve generalization and reduce overfitting.
\subsection{Mamba Block and U-Mamba Network}
\label{sec:Mamba:UMamba}

The U-Mamba network integrates the strengths of the Mamba block~\cite{gu2023mamba} and U-Net~\cite{ronneberger2015u} to enhance global context understanding and improve the accuracy of medical image segmentation. The Mamba block, designed for Selective Structured State Space Sequence Models (S6), efficiently handles long-range dependencies and sequential information, making it highly suitable for medical image segmentation tasks.

The State Space Model (SSM)~\cite{gu2021combining} describes a system in terms of its internal state and observations over time, effectively modeling sequences through underlying states. Given the input state vector \(\mathbf{x}_t\), control input \(\mathbf{u}_t\), process noise \(\mathbf{w}_t\), state transition matrix \(\mathbf{A}\), and control input matrix \(\mathbf{B}\), the basic form is represented by:
\begin{equation}
    \begin{aligned}
        \mathbf{x}_{t+1} = \mathbf{A}\mathbf{x}_t + \mathbf{B}\mathbf{u}_t + \mathbf{w}_t.
    \end{aligned}
    \label{eq:SSMs}
\end{equation}
For the observation matrix \(\mathbf{C}\), feedthrough matrix \(\mathbf{D}\), and observation noise \(\mathbf{v}_t\), the observation \(\mathbf{y}_t\) is then given by:
\begin{equation}
    \begin{aligned}
        \mathbf{y}_t = \mathbf{C}\mathbf{x}_t + \mathbf{D}\mathbf{u}_t + \mathbf{v}_t.
    \end{aligned}
    \label{eq:observation_at_timet}
\end{equation}

The S6 architecture enhances traditional state space models with selective attention mechanisms and structured parameterization. Specifically, the selective attention mechanism can be expressed as:
\begin{equation}
    \begin{aligned}
        \mathbf{a}_t = \text{softmax}(\mathbf{Q} \mathbf{K}^T / \sqrt{d_k}) \mathbf{V},
    \end{aligned}
    \label{eq:attention}
\end{equation}
where \(\mathbf{Q}\), \(\mathbf{K}\), and \(\mathbf{V}\) are the query, key, and value matrices derived from the state vector \(\mathbf{x}_t\); \(d_k\) is the dimension of the key vectors. This selective attention mechanism focuses on relevant input sequence parts and captures complex dependencies.

Integrating S6 within the Mamba block is crucial for sequential medical image processing, such as cardiac MRI segmentation, where capturing temporal dynamics and structure is essential. Our approach focuses exclusively on per-image segmentation tasks, treating each image as an independent entity, which means the state transition and observation matrices (\(\mathbf{A}\), \(\mathbf{C}\), {\em etc.}) are applied to individual images.

U-Mamba leverages Mamba's linear scaling advantage to improve CNNs' long-range dependency modeling without the high computational burden of self-attention mechanisms used in Transformers like ViT \cite{vaswani2017attention} and SwinTransformer~\cite{liu2021swin}. Fig.~\ref{fig:Overview_proposed_model}  shows the U-Mamba block, which consists of two sequential residual blocks followed by a Mamba block.
Each block includes convolutional layers, Instance Normalization, and Leaky ReLU activation. Image features are flattened, transposed, normalized, and processed through Mamba blocks with two parallel branches: one with an SSM layer and one without. These features are merged via the Hadamard product, projected back, and transposed to their original shape.
The complete U-Mamba network architecture features an encoder with these blocks to capture local features and long-range dependencies, and a decoder composed of residual blocks and transposed convolutions for detailed local information and resolution recovery. Skip connections link hierarchical features from the encoder to the decoder, and the final decoder feature is passed through a 1$\times$1$\times$1 convolutional layer and a Softmax layer to produce the final segmentation probability map.

\subsection{Uncertainty-aware Loss}
\label{sec:uncertainty-aware:loss}

Integrating uncertainty into loss functions involves weighting different elements of a loss function according to the estimated uncertainty for each data point~\cite{zhao2020uncertainty,zhao2019uncertainty}. This approach allows the model to prioritize learning from reliable instances and minimize the impact of potentially erroneous or ambiguous data.
Kendall and Gal introduced the use of homoscedastic and heteroscedastic uncertainty to adjust loss functions~\cite{kendall2017uncertainties}. 
Homoscedastic uncertainty is constant across all data points, while heteroscedastic uncertainty varies between instances. This technique enables the model to adapt its learning process based on these uncertainties, enhancing resilience and precision by focusing on confident predictions and reducing the influence of ambiguous ones. As a result, it optimizes the training process across various datasets and improves overall performance~\cite{hu2023umednerf,wang2024neural,peng2024uncertainty}.

To enhance segmentation accuracy, we implement an uncertainty-aware loss function that combines the following region-based, distribution-based, and pixel-based losses, leveraging their complementary strengths.

\begin{enumerate}[leftmargin=12pt] \itemsep -.1em
    \item {\bf Dice Coefficient (DC) loss}: This region-based metric emphasizes the overlap between predicted and ground truth areas, preserving the accuracy of the shape and boundaries of segmented regions.
    \item {\bf Cross-Entropy (CE) loss}: This distribution-based loss ensures accurate categorization of individual pixels, improving classification precision.
    \item {\bf Focal loss}: This pixel-level loss addresses class imbalance by assigning greater importance to difficult-to-classify instances, enhancing the model's capability to handle complex scenarios~\cite{hu2020learning, hu2022sum, hu2023rank}.
\end{enumerate}

Given the predicted segmentation and the respective ground-truth mask, let \(p_i\) denote the predicted probability and \(g_i\) denote the ground truth label. The Dice Similarity Coefficient (DSC) measures the overlap between the predicted segmentation and the ground truth. It is defined as:
\begin{equation}
    \begin{aligned}
        DSC = \frac{2 \sum_i p_i g_i}{\sum_i p_i + \sum_i g_i}
    \end{aligned}
    \label{eq:DSC}
\end{equation}
The DSC ranges from 0 to 1, where a DSC of 1 indicates perfect overlap between the prediction and the ground truth, and a DSC of 0 indicates no overlap.

To use this measure in a loss function for training segmentation models, we define the Dice Coefficient (DC) loss as:
\begin{equation}
    \begin{aligned}
        \mathcal{L}_{DC} = 1 - DSC = 1 - \frac{2 \sum_i p_i g_i}{\sum_i p_i + \sum_i g_i}
    \end{aligned}
    \label{eq:DC_loss}
\end{equation}
The DC loss aims to minimize the difference between the predicted segmentation and the ground truth by maximizing the DSC. By minimizing the DC loss, the model is trained to produce segmentations that have a higher overlap with the ground truth, thus improving segmentation accuracy.

The Cross-Entropy (CE) loss is defined following the standard entropy calculation:
\begin{equation}
    \begin{aligned}
        \mathcal{L}_{CE} = -\sum_i g_i \log(p_i)
    \end{aligned}
    \label{eq:CE_loss}
\end{equation}

The Focal loss is designed to address class imbalance by focusing more on hard-to-classify samples:
\begin{equation}
\begin{aligned}
    \mathcal{L}_{focal} = -\sum_i (1 - p_i)^\gamma g_i \log(p_i)
\end{aligned}
\label{eq:focal_loss}
\end{equation}
where \(\gamma\) is a focusing parameter default to 2.

We define the uncertainty-aware loss $\mathcal{L}_{UA}$ by combining these losses into an uncertainty-aware framework:
\begin{equation}
    \begin{aligned}
        \mathcal{L}_{UA} = \sum_{m=1}^{M} \left( \frac{1}{2\sigma_m^2} \mathcal{L}_m + \log(1 + \sigma_m^2) \right)
    \end{aligned}
    \label{eq:uncertainty-aware_loss}
\end{equation}

where \( M \) is the number of individual loss components, \( \mathcal{L}_m \) represents each individual loss component (such as DC, CE, and focal loss), and \( \sigma_m \) are learnable parameters that adjust the contribution of each loss component based on uncertainty. These learnable parameters are optimized during the training process to minimize the overall loss.

Introducing uncertainty into the loss computation allows the model to dynamically assign weights to each loss component. This approach effectively balances global and local accuracy while mitigating the impact of class imbalance. Consequently, the model becomes more resilient to noisy or ambiguous data, leading to overall improved segmentation performance.

\subsection{Sharpness-Aware Minimization Optimization}
\label{sec:SAM:opt}

Our approach involves utilizing Sharpness-Aware Minimization (SAM) optimization to improve the performance of the U-Mamba model in segmenting cardiac MRI images such as those on the ACDC dataset~\cite{bernard2018deep}. SAM optimization flattens the loss landscape, thereby improving the model's generalization potential. Conventional optimization techniques aim to find the  lowest points in the loss landscape, but these points can be steep and lead to poor generalization on new data. In contrast, SAM identifiers smoother minima, which are regions in the parameter space where the model's performance is more consistent and less susceptible to disturbances.

The motivation for utilizing SAM stems from its ability to mitigate overfitting, a common challenge in medical image segmentation. Traditional optimization techniques often result in narrow valleys in the loss landscape, leading to suboptimal performance on new data. SAM, however, targets flatter minima, which are associated with better generalization performance. This is particularly beneficial when working with complex and diverse datasets like ACDC, where the variability in cardiac MRI images can lead to overfitting if not properly managed.

The SAM optimization involves an iterative two-step process. First, for every mini-batch, the parameters are adjusted in the direction that maximizes the loss. Next, the model parameters are updated to minimize the highest altered loss. The purpose of this perturbation is to find model parameters that reside in flatter regions of the loss landscape, which typically correspond to better generalization and robustness to small changes in the input data.
Specifically, let \(\theta\) represent the model parameters, \(\mathcal{L}\) denote the loss function, and \(\mathcal{D}\) denote the training dataset. Let \(\epsilon\) denote the perturbation, and \(\rho\) control the size of the neighborhood around \(\theta\). The perturbation \(\epsilon\) is added to the parameters \(\theta\) to explore the loss landscape in a local neighborhood defined by the norm constraint \(\|\epsilon\|_2 \leq \rho\).

The SAM optimization is mathematically formulated as:
\begin{equation}
\begin{aligned}
    \theta^* = \arg \min_{\theta} \max_{\epsilon: \|\epsilon\|_2 \leq \rho} \mathcal{L}(\theta + \epsilon; \mathcal{D}).
\end{aligned}
\label{eq:SAM_optimization}
\end{equation}

The first step involves finding the perturbation \(\epsilon\) within the  \(\|\epsilon\|_2 \leq \rho\) that maximizes the loss. This step identifies the worst-case direction in the local neighborhood of \(\theta\), ensuring that the model parameters are adjusted towards regions of the loss landscape that are not steep.
In the second step, the model parameters \(\theta\) are updated to minimize the loss at this worst-case perturbed location, making the parameters robust to perturbations.
By iteratively applying these two steps, SAM ensures the optimizer seeks parameter configurations robust to perturbations, improving generalization and performance. This approach helps the model to find flatter minima in the loss landscape with better generalization~\cite{foret2021sharpness}.
\begin{figure*}[t]
    \centerline{
    \includegraphics[width=0.99\textwidth]{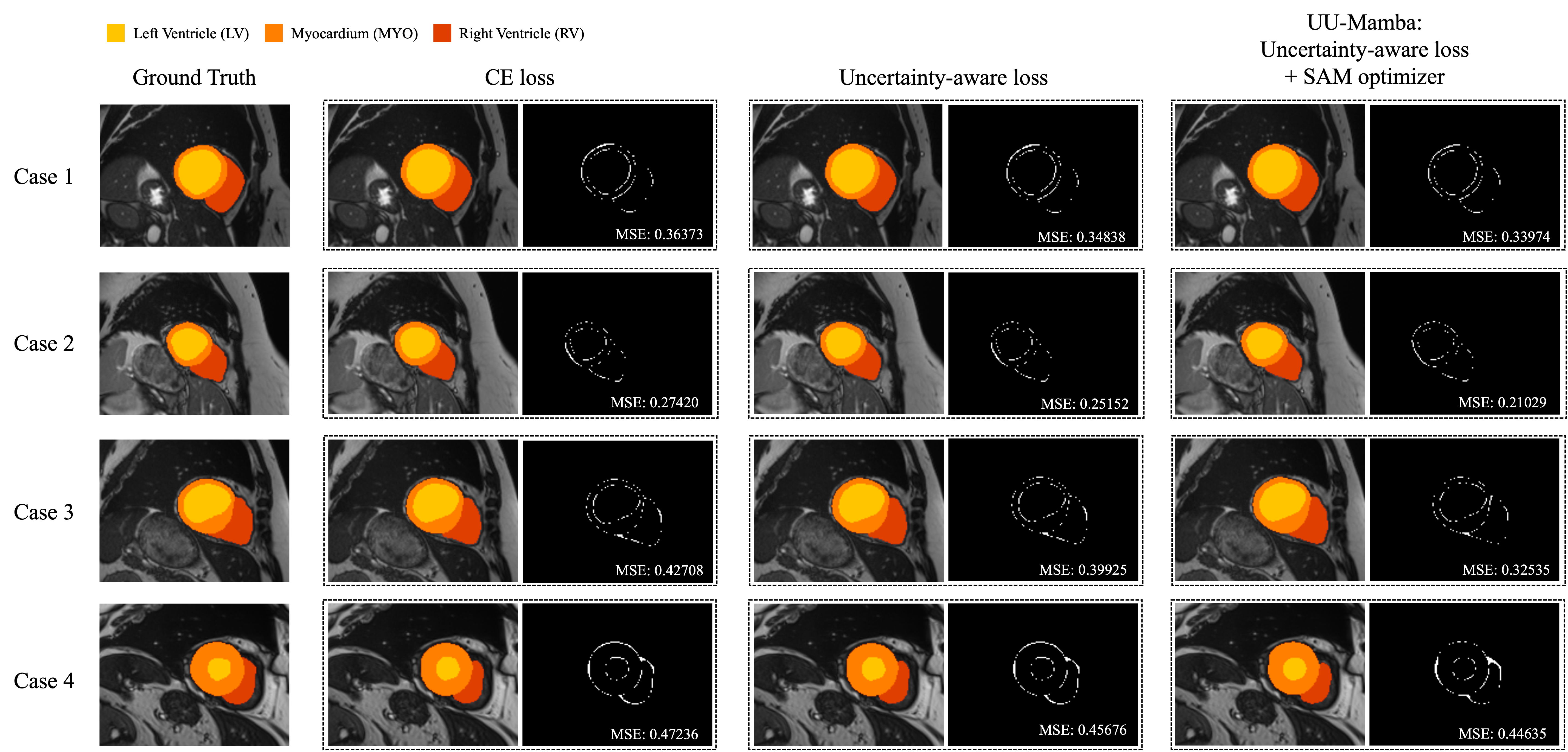}
    \vspace{-2mm}
    }  
    \caption{Segmentation results for various methods on sample images from the ACDC dataset. The Mean Squared Error (MSE) between the output segmentation and the ground truth is shown for each method.}
    \vspace{-3mm}
    \label{fig:segmentation_results}
\end{figure*}
\section{Experiments}
\subsection{Experimental Settings}

\subsubsection{Datasets}

The Automated Cardiac Diagnosis Challenge (ACDC) dataset~\cite{bernard2018deep} is a widely used benchmark in the field of medical image analysis, particularly for cardiac MRI segmentation tasks. The dataset comprises MRI scans of 150 patients, each with multiple slices, aggregating to a total of 300 images and 2,978 slices. Patients in the dataset are distributed evenly across five groups, deliberated by distinct cardiac pathologies: {\em normal subjects}, {\em myocardial infarction}, {\em dilated cardiomyopathy}, {\em hypertrophic cardiomyopathy}, and {\em abnormal right ventricle}.
The dataset features short-axis cardiac MRI images, which provide comprehensive heart coverage. Ground truth labels for the left ventricle (LV), right ventricle (RV), and myocardium (MYO) are provided for each image, facilitating the development and evaluation of segmentation algorithms.
Variability is present in both image spacing and size across different dimensions within the ACDC dataset.

\subsubsection{Evaluation Metrics}

Following the evaluation protocol outlined in \cite{maier-hein2022metrics}, we utilize the Dice Similarity Coefficient (DSC) as our primary metric for segmentation evaluation. The DSC quantifies the overlap between the predicted segmentation and the ground truth mask, offering a robust assessment of the model's accuracy in delineating cardiac structures. The DSC is defined in Eq.~\eqref{eq:DSC}.
Additionally, we use the Mean Squared Error (MSE) to measure the average squared difference between the predicted probabilities and the ground truth labels. Let \(N\) is the number of testing images, \(H\) and \(W\) are the height and width of the images, \(p_{ij}^n\) is the predicted probability at pixel \((i, j)\) for the \(n\)-th image, and \(g_{ij}^n\) is the ground truth label at pixel \((i, j)\) for the \(n\)-th image. The MSE is calculated as follows:
\begin{equation}
\begin{aligned}
    \text{MSE} = \frac{1}{N} \sum_{n=1}^{N} \frac{1}{HW} \sum_{i=1}^{H} \sum_{j=1}^{W} (p_{ij}^n - g_{ij}^n)^2
\end{aligned}
\label{eq:MSE}
\end{equation}
MSE provides a complementary evaluation that helps in understanding performance in terms of pixel-wise precision.

\subsubsection{Implementation Details}

The experiments were conducted using the PyTorch framework and two NVIDIA A100 Tensor Core GPUs for training. During training, a patch size of [20, 256, 224] and a batch size of 4 were employed. The network configuration comprises 6 stages, with the number of pooling operations per axis set to [2, 5, 5]. An initial learning rate of $5 \times 10^{-3}$ was utilized, and the training proceeded for 500 epochs. In the SAM optimization, the hyperparameter $\rho$ controlling the perturbation in Eq.~\eqref{eq:SAM_optimization} was set to 0.05. The focusing parameter $\gamma$ in the focal loss in Eq.~\eqref{eq:focal_loss} was set to 2. The parameter \( M \) of the uncertainty-aware loss in Eq.~\eqref{eq:uncertainty-aware_loss} was set to 3, incorporating DC loss, CE loss, and focal loss.

\begin{table*}[t]
\centerline{
    \begin{tabular}{c|c|c|c|c|c}
        \hline
        Method                                 & Avg. DSC \% $\uparrow$ & RV DSC \% $\uparrow$ & Myo DSC \% $\uparrow$ & LV DSC \% $\uparrow$ & Avg. MSE \% $\downarrow$ \\ \hline
        CE loss                                & 92.263                 & 91.81                & 90.31                 & 94.67                & 0.3843                   \\ \hline
        Uncertainty-aware loss (CE, DC, Focal) & 92.602                 & 92.36                & 90.51                 & 94.94                & 0.3640                   \\ \hline
        Uncertainty-aware loss + SAM (Ours)    & \textbf{92.787}        & \textbf{92.41}       & \textbf{90.90}        & \textbf{95.04}       & \textbf{0.3304}          \\ \hline
    \end{tabular}
}
\caption{Ablation study on various configurations with U-Manmba backbone: (1) using only the traditional Cross-Entropy (CE) loss, (2) using the uncertainty-aware Loss without the SAM optimizer, and (3) the proposed UU-Mamba model. The evaluation metric is DSC (\%) and MSE.
}
\label{tab:Results_ablation_study}
\vspace{-3mm}
\end{table*}

\subsection{Experimental Results}
We compare UU-Mamba against five leading segmentation models: TransUNet~\cite{Chen2021TransUNet}, Swin-Unet~\cite{Cao2021SwinUnet}, nnUNet~\cite{Isensee2021nnUNet}, nnFormer~\cite{Zhou2021nnFormer}, and U-Mamba~\cite{ma2024umamba}. TransUNet and Swin-Unet are Transformer-based networks, while nnUNet and nnFormer incorporate CNN-based architectures. U-Mamba is a hybrid architecture that integrates elements from both Transformer-based networks and CNN-based architectures.

We conducted a quantitative evaluation of our UU-Mamba against the five other 3D heart area segmentation models on the ACDC dataset, using DSC as the evaluation metric.
Fig.~\ref{fig:segmentation_results} shows the segmentation results for the methods in comparison on a few images from the ACDC dataset. 
Table~\ref{tab:Results_comparison_diff_models} reports the DSC scores for each model at three heart regions: the right ventricle (RV), myocardium (Myo), and left ventricle (LV), as well as the average DSC scores of all regions. The best performing scores are marked in bold.

\begin{table}[t]
\centerline{
\begin{tabular}{c|c|c|c|c}
        \hline
        Method                              & Average        & RV $\uparrow$    & Myo $\uparrow$   & LV $\uparrow$    \\ \hline
        TransUNet \cite{Chen2021TransUNet}  & 89.71          & 88.86          & 84.53          & 95.73          \\ \hline 
        Swin-Unet \cite{Cao2021SwinUnet}    & 90.00          & 88.55          & 85.62          & \textbf{95.83} \\ \hline
        nnUNet    \cite{Isensee2021nnUNet}  & 91.61          & 90.24          & 89.24          & 95.36          \\ \hline
        nnFormer  \cite{Zhou2021nnFormer}   & 92.06          & 90.94          & 89.58          & 95.65          \\ \hline
        U-Mamba   \cite{ma2024umamba}       & 92.22          & 91.83          & 90.22          & 94.54          \\ \hline
        UU-Mamba (Ours)                     & \textbf{92.79} & \textbf{92.41} & \textbf{90.90} & 95.04          \\ \hline
    \end{tabular}
}
\caption{
Performance comparison of our UU-Mamba with leading medical image segmentation methods on the ACDC dataset. The evaluation metric is DSC (\%).}
\label{tab:Results_comparison_diff_models}
\vspace{-3mm}
\end{table}

Our UU-Mamba model demonstrates the highest overall performance, achieving an average DSC of 92.79\%. The scores for each individual region are as follows: 92.41\% for RV, 90.90\% for Myo, and 95.04\% for LV. Our method surpasses all other models in accurately segmenting different heart components in the right ventricle and myocardium regions, showcasing its flexibility and efficacy. In Table~\ref{tab:Results_comparison_diff_models}, the slight decrease in DSC for the left ventricle, compared to other models, is mitigated by the higher scores exhibited in other regions and the highest average DSC.

In comparison, TransUNet achieves an average DSC of 89.71\%. While it performs well overall, it shows a lower DSC for Myo compared to other regions. Swin-Unet achieves an average DSC of 90.00\%, exhibiting the highest DSC for the LV but a lower DSC for RV, compared to other regions. The nnUNet model achieves an average DSC of 91.61\%, demonstrating significant improvement in DSC for Myo and overall performance. The nnFormer model achieves an average DSC of 92.06\%, offering the best performance among the four comparing models across all regions, particularly the Myo. The U-Mamba method further improves upon these results with an average DSC of 92.22\%, showing notable performance gains in the RV and Myo regions.

The quantitative evaluation highlights the superior performance of our UU-Mamba compared to existing models. With the highest average DSC and exceptional segmentation of the right ventricle and myocardium, our method demonstrates its potential for advancing the accuracy of 3D heart segmentation in medical imaging. The improvements over other models like U-Mamba, nnFormer and nnUNet highlight the effectiveness of our approach in leveraging both global and local features through innovative loss functions and optimization techniques.
\subsection{Ablation Study}

We perform ablation study of our UU-Mamba model, by taking out the SAM optimization, and incorporating only the cross-entropy (CE) loss or the uncertainty-aware loss, and compare the respective segmentation performance on the ACDC dataset. Table~\ref{tab:Results_ablation_study} summarizes these results. The results demonstrate that incorporating uncertainty-aware loss and the SAM optimization indeed improves segmentation accuracy.

Training using only the CE loss yields 92.263\% DSC. In comparison, training using the uncertainty-aware loss (which incorporates the DC loss, CE loss, and focused loss), enhances DSC to 92.602\%. This result shows the capability of the uncertainty-aware loss to increase the robustness and accuracy of models by prioritizing confident predictions and minimizing the impact of uncertain ones. It effectively addresses the limitations of using a single loss function by balancing various aspects of the segmentation task.

Incorporating the SAM optimization into the uncertainty-aware loss yields 92.787\% DSC, the highest among the examined approaches. SAM  improves segmentation accuracy by generating clearer and more accurate borders, guiding the optimization process toward flat minima in the loss landscape. This leads to better generalization across diverse data samples. The increase in DSC from 92.263\% to 92.787\% highlights the effectiveness of SAM in enhancing segmentation results and more precisely identifying heart components.

\subsection{Robustness Analysis}

We perform experiments to evaluate each method on the Mean Squared Error (MSE) of the DSC scores to assess their robustness quantitatively. The MSE is calculated as shown in Eq.~\eqref{eq:MSE}. Results are shown in the Table~\ref{tab:Results_ablation_study}. These results show that the uncertainty-aware loss reduces the MSE compared to the standard CE loss, reflecting its ability to better address the variability and uncertainty in the data. The inclusion of SAM optimization significantly decreases the MSE, achieving the lowest error value. This reduction in MSE highlights SAM's role in minimizing errors and producing more accurate segmentation maps.
Fig.~\ref{fig:segmentation_results} shows the MSE between the output segmentation and the ground truth  for each method, providing a visual comparison of the segmentation quality.

\section{Conclusion}

We introduce a new model, UU-Mamba, designed for 3D heart segmentation in cardiac MRI images. This model integrates the U-Mamba architecture with an uncertainty-aware loss function and the SAM optimizer. Our method improves biological image segmentation by improving generalization and boundary accuracy. 
The uncertainty-aware loss combines region-based, distribution-based, and pixel-based losses to improve segmentation performance by harmonizing tasks and prioritizing confident predictions. Meanwhile, the SAM optimizer guides the model toward flat minima in the loss landscape, enhancing resilience and reducing overfitting, ultimately resulting in more accurate segmentation.
Comparative analysis against five leading models demonstrates the superiority of UU-Mamba, achieving a DSC of 92.787\%, signifying both accuracy and robustness in segmentation performance.

{\bf Future work} will encompass the exploration of additional data augmentation techniques, the investigation of alternative uncertainty modeling methods, and the validation of the model on larger and more diverse datasets. Our overarching aim is to advance automated medical imaging by continuously refining and broadening the UU-Mamba approach.

\noindent
\textbf{Acknowledgments.}
This work is supported by the University at Albany Start-up Grant. The authors appreciate the computational resource provided by the University at Albany -- SUNY. 

\small
\bibliographystyle{plain}
\bibliography{main}

\begin{thebibliography}{10}

\bibitem{azad2023}
Reza Azad, Moein Heidary, Kadir Yilmaz, Michael Hüttemann, Sanaz Karimijafarbigloo, Yuli Wu, Anke Schmeink, and Dorit Merhof.
\newblock {Loss Functions in the Era of Semantic Segmentation: A Survey and Outlook}.
\newblock {\em arXiv preprint arXiv:2312.05391}, 2023.

\bibitem{bernard2018deep}
Olivier Bernard, Antoine Lalande, Caian Zotti, Filip Cervenansky, Xin Yang, Pheng-Ann Heng, Irem Cetin, Karim Lekadir, Oscar Camara, Miguel Angel~Gonzalez Ballester, et~al.
\newblock {Deep Learning Techniques for Automatic {MRI} Cardiac Multi-Structures Segmentation and Diagnosis: Is the Problem Solved?}
\newblock {\em IEEE Transactions on Medical Imaging}, 37(11):2514--2525, 2018.

\bibitem{Cao2021SwinUnet}
Hu~Cao, Yueyi Wang, Joy Chen, Dong Jiang, Xiaopeng Zhang, Qi~Tian, and Manning Wang.
\newblock {{Swin-Unet}: Unet-like pure transformer for medical image segmentation}.
\newblock {\em arXiv preprint arXiv:2105.05537}, 2021.

\bibitem{chen2020deep}
Cheng Chen, Chen Qin, Hongwei Qiu, Giacomo Tarroni, Jin Duan, Wenjia Bai, and Daniel Rueckert.
\newblock {Deep Learning for Cardiac Image Segmentation: A Review}.
\newblock {\em Frontiers in Cardiovascular Medicine}, 7:25, 2020.

\bibitem{Chen2021TransUNet}
Jieneng Chen, Yongyi Lu, Qihang Yu, Xiangde Luo, Ehsan Adeli, Yun Wang, Le~Lu, Yuyin Zhou, Yefeng Wang, and Alan Yuille.
\newblock {{TransUNet}: Transformers make strong encoders for medical image segmentation}.
\newblock {\em arXiv preprint arXiv:2102.04306}, 2021.

\bibitem{cipolla2018}
Roberto Cipolla, Yarin Gal, and Alex Kendall.
\newblock {Multi-task Learning Using Uncertainty to Weigh Losses for Scene Geometry and Semantics}.
\newblock In {\em 2018 IEEE/CVF Conference on Computer Vision and Pattern Recognition}, pages 7482--7491, Salt Lake City, UT, USA, 2018. IEEE.

\bibitem{fahmy2019}
Ahmed~S. Fahmy, Hossam El-Rewaidy, Maryam Nezafat, Shiro Nakamori, and Reza Nezafat.
\newblock Automated analysis of cardiovascular magnetic resonance myocardial native t1 mapping images using fully convolutional neural networks.
\newblock {\em Journal of Cardiovascular Magnetic Resonance}, 21(1):7, 2019.

\bibitem{foret2021sharpness}
Pierre Foret, Ariel Kleiner, Hossein Mobahi, and Behnam Neyshabur.
\newblock {Sharpness-Aware Minimization for Efficiently Improving Generalization}.
\newblock In {\em Proceedings of the 8th International Conference on Learning Representations (ICLR)}, 2021.

\bibitem{gu2023mamba}
Albert Gu and Tri Dao.
\newblock {{Mamba}: Linear-Time Sequence Modeling with Selective State Spaces}.
\newblock {\em arXiv preprint arXiv:2312.00752}, 2023.

\bibitem{gu2021combining}
Albert Gu, Isys Johnson, Karan Goel, Khaled Saab, Tri Dao, Atri Rudra, and Christopher R{\'e}.
\newblock {Combining Recurrent, Convolutional, and Continuous-time Models with Linear State-Space Layers}.
\newblock {\em Advances in Neural Information Processing Systems}, 34, 2021.

\bibitem{hu2023umednerf}
Jing Hu, Qinrui Fan, Shu Hu, Siwei Lyu, Xi~Wu, and Xin Wang.
\newblock {{UMedNeRF}: Uncertainty-aware Single View Volumetric Rendering for Medical Neural Radiance Fields}.
\newblock {\em IEEE International Symposium on Biomedical Imaging}, 2024.

\bibitem{hu2023rank}
Shu Hu, Xin Wang, and Siwei Lyu.
\newblock {Rank-based Decomposable Losses in Machine Learning: A Survey}.
\newblock {\em IEEE Transactions on Pattern Analysis and Machine Intelligence}, 2023.

\bibitem{hu2020learning}
Shu Hu, Yiming Ying, Xin Wang, and Siwei Lyu.
\newblock {Learning by Minimizing the Sum of Ranked Range}.
\newblock {\em Advances in Neural Information Processing Systems}, 33:21013--21023, 2020.

\bibitem{hu2022sum}
Shu Hu, Yiming Ying, Xin Wang, and Siwei Lyu.
\newblock {Sum of Ranked Range Loss for Supervised Learning}.
\newblock {\em Journal of Machine Learning Research}, 23(112):1--44, 2022.

\bibitem{Isensee2021nnUNet}
Fabian Isensee, Paul~F. Jaeger, Simon A.~A. Kohl, Jens Petersen, and Klaus~H. Maier-Hein.
\newblock {{nnU-Net}: a self-configuring method for deep learning-based biomedical image segmentation}.
\newblock {\em Nature Methods}, 18(2):203--211, 2021.

\bibitem{kendall2017uncertainties}
Alex Kendall and Yarin Gal.
\newblock {What Uncertainties Do We Need in Bayesian Deep Learning for Computer Vision?}
\newblock In {\em Proceedings of the 31st International Conference on Neural Information Processing Systems}, NIPS'17, page 5580–5590, Red Hook, NY, USA, 2017. Curran Associates Inc.

\bibitem{lin2024preserving}
Li~Lin, Xinan He, Yan Ju, Xin Wang, Feng Ding, and Shu Hu.
\newblock {Preserving Fairness Generalization in Deepfake Detection}.
\newblock {\em CVPR}, 2024.

\bibitem{lin2024robust2}
Li~Lin, Yamini~Sri Krubha, Zhenhuan Yang, Cheng Ren, Xin Wang, and Shu Hu.
\newblock {Robust {COVID-19} Detection in {CT} Images with {CLIP}}.
\newblock {\em MIPR}, 2024.

\bibitem{lin2024robust}
Li~Lin, Sarah Papabathini, Xin Wang, and Shu Hu.
\newblock {Robust Light-Weight Facial Affective Behavior Recognition with {CLIP}}.
\newblock {\em MIPR}, 2024.

\bibitem{litjens2017survey}
Geert Litjens, Thijs Kooi, Babak~Ehteshami Bejnordi, Arnaud Arindra~Adiyoso Setio, Francesco Ciompi, Mohsen Ghafoorian, Jeroen~AWM van~der Laak, Bram van Ginneken, and Clara~I S{\'a}nchez.
\newblock {A survey on deep learning in medical image analysis}.
\newblock {\em Medical image analysis}, 42:60--88, 2017.

\bibitem{liu2021swin}
Ze~Liu, Yutong Lin, Yue Cao, Han Hu, Yixuan Wei, Zheng Zhang, Stephen Lin, and Baining Guo.
\newblock {{Swin Transformer}: Hierarchical Vision Transformer using Shifted Windows}.
\newblock {\em Proceedings of the IEEE/CVF International Conference on Computer Vision}, pages 10012--10022, 2021.

\bibitem{ma2024umamba}
Jun Ma, Feifei Li, and Bo~Wang.
\newblock {{U-Mamba}: Enhancing Long-range Dependency for Biomedical Image Segmentation}.
\newblock {\em arXiv preprint arXiv:2401.04722}, 2024.

\bibitem{maier2019gentle}
Andreas~K Maier, Christoph Syben, Tobias Lasser, and Christian Riess.
\newblock A gentle introduction to deep learning in medical image processing.
\newblock {\em Zeitschrift f{\"u}r Medizinische Physik}, 29(2):86--101, 2019.

\bibitem{maier-hein2022metrics}
Lena Maier-Hein, Annika Reinke, Patrick Godau, Minu~D. Tizabi, Florian Buettner, Evangelia Christodoulou, Ben Glocker, Fabian Isensee, Jens Kleesiek, Michal Kozubek, et~al.
\newblock {Metrics reloaded: recommendations for image analysis validation}.
\newblock {\em Nature Methods}, 21(2):195–212, feb 2024.

\bibitem{Mariam2024}
Iqra Mariam, Xiaorong Xue, and Kaleb Gadson.
\newblock {A Retinal Vessel Segmentation Method Based on the Sharpness-Aware Minimization Model}.
\newblock {\em Sensors}, 24(13), 2024.

\bibitem{peng2024uncertainty}
Yicui Peng, Hao Chen, Chingsheng Lin, Guo Huang, Jinrong Hu, Hui Guo, Bin Kong, Shu Hu, Xi~Wu, and Xin Wang.
\newblock {{Uncertainty-Aware} Explainable Recommendation with Large Language Models}.
\newblock {\em IJCNN}, 2024.

\bibitem{petitjean2011review}
Caroline Petitjean and Jean-Nicolas Dacher.
\newblock {A review of segmentation methods in short axis cardiac {MR} images}.
\newblock {\em Medical image analysis}, 15(2):169--184, 2011.

\bibitem{ronneberger2015u}
Olaf Ronneberger, Philipp Fischer, and Thomas Brox.
\newblock {{U-Net}: Convolutional networks for biomedical image segmentation}.
\newblock {\em International Conference on Medical image computing and computer-assisted intervention}, pages 234--241, 2015.

\bibitem{lin2024robust3}
Santosh, Li~Lin, Irene Amerini, Xin Wang, and Shu Hu.
\newblock {Robust {CLIP}-Based Detector for Exposing Diffusion Model-Generated Images}.
\newblock {\em arXiv preprint arXiv:2404.12908}, 2024.

\bibitem{vaswani2017attention}
Ashish Vaswani, Noam Shazeer, Niki Parmar, Jakob Uszkoreit, Llion Jones, Aidan~N Gomez, Łukasz Kaiser, and Illia Polosukhin.
\newblock {Attention is all you need}.
\newblock {\em Advances in neural information processing systems}, 30, 2017.

\bibitem{wang2024neural}
Xin Wang, Shu Hu, Heng Fan, Hongtu Zhu, and Xin Li.
\newblock {Neural Radiance Fields in Medical Imaging: Challenges and Next Steps}.
\newblock {\em arXiv preprint arXiv:2402.17797}, 2024.

\bibitem{wang2024artificial}
Xin Wang and Hongtu Zhu.
\newblock {Artificial Intelligence in Image-based Cardiovascular Disease Analysis: A Comprehensive Survey and Future Outlook}.
\newblock {\em arXiv:2402.03394}, 2024.

\bibitem{wang2024weakmambaunet}
Ziyang Wang and Chao Ma.
\newblock {{Weak-Mamba-UNet}: Visual Mamba Makes {CNN} and {ViT} Work Better for Scribble-based Medical Image Segmentation}.
\newblock {\em arXiv preprint arXiv:2402.10887}, 2024.

\bibitem{wang2024mambaunet}
Ziyang Wang, Jian-Qing Zheng, Yichi Zhang, Ge~Cui, and Lei Li.
\newblock {{Mamba-UNet}: {UNet-Like} Pure Visual Mamba for Medical Image Segmentation}.
\newblock {\em arXiv preprint arXiv:2402.05079}, 2024.

\bibitem{zhao2020uncertainty}
Xujiang Zhao, Feng Chen, Shu Hu, and Jin-Hee Cho.
\newblock {Uncertainty Aware Semi-Supervised Learning on Graph Data}.
\newblock {\em Advances in Neural Information Processing Systems}, 33:12827--12836, 2020.

\bibitem{zhao2019uncertainty}
Xujiang Zhao, Shu Hu, Jin-Hee Cho, and Feng Chen.
\newblock {Uncertainty-based Decision Making Using Deep Reinforcement Learning}.
\newblock In {\em 2019 22th International Conference on Information Fusion (FUSION)}, pages 1--8. IEEE, 2019.

\bibitem{Zhou2021nnFormer}
Huiyu Zhou, Yuyin Yang, Kai Chen, Jie Huang, Yucheng Pan, Yingwei Wang, Xin Xie, Liansheng Yang, and Huazhong Shu.
\newblock {{nnFormer}: Volumetric Medical Image Segmentation via a {3D} Transformer}.
\newblock {\em arXiv preprint arXiv:2109.03201}, 2021.

\end{thebibliography}

\end{document}